# Calculation of Local Pressure Tensors in Systems with Many-Body Interactions[‡]

Hendrik Heinz*[#], Wolfgang Paul[#], and Kurt Binder[#]

[#] Institute of Physics, Johannes-Gutenberg-University of Mainz, D-55099 Mainz, Germany

* Department of Materials, Institute of Polymers, ETH Zurich, Universitaetsstr. 6, CH-8092 Zurich, Switzerland; correspondence to hendrik.heinz@wright.edu

**Abstract**

Local pressures are important in the calculation of interface tensions and in analyzing micromechanical behaviour. The calculation of local pressures in computer simulations has been limited to systems with pairwise interactions between the particles, which is not sufficient for chemically detailed systems with many-body potentials such as angles and torsions. We introduce a method to calculate local pressures in systems with $n$-body interactions ($n = 2, 3, 4, \ldots$) based on a micromechanical definition of the pressure tensor. The local pressure consists of a kinetic contribution from the linear momentum of the particles and an internal contribution from dissected many-body interactions by infinitesimal areas. To define dissection by a small area, respective $n$-body interactions are divided into two geometric centers, effectively reducing them to two-body interactions. Consistency with hydrodynamics-derived formulas for systems with two-body interactions [J. H. Irving and J. G. Kirkwood, J. Chem. Phys. 18, 817 (1950)], for average cross-sectional pressures [B. D. Todd, D. J. Evans, and P. J. Daivis, Phys. Rev. E 52, 1627 (1995)], and for volume averaged pressures (virial formula) is shown. As a simple numerical example, we discuss liquid propane in a cubic box. Local, cross-sectional, and volume-averaged pressures as well as relative contributions from 2-body and 3-body forces are analyzed with the proposed method, showing full numerical equivalence with the existing approaches. The method allows computing local pressures in the presence of many-body interactions in atomistic simulations of complex materials and biological systems.

# 1. INTRODUCTION

Local pressures are important for the calculation of interface tensions and the analysis of mechanical responses to strain, heat, photoexcitation, and phase transformations [1,2]. The study of such processes is broadly relevant in condensed matter physics, chemistry, and materials science. For example, the analysis of local pressures in fluids [3,4], polymers [5-7], at surfaces [8,9], as well as during conformational changes in proteins [10,11] contributed to the understanding of interfaces and secondary molecular structure.

Calculations of local pressures in computer simulations became feasible through the seminal work of Irving and Kirkwood on the Navier-Stokes equation of hydrodynamics [12]. Irving and Kirkwood described the calculation of local pressure tensors in the presence of pairwise interactions between the particles. However, the limitation to two-body interactions poses a problem for the calculation of local pressures in computer simulations of biochemical or materials systems [13]. Many-body potentials such as angle bending and torsion potentials are often essential to understand the local structure and dynamics [13], which necessitates a method to calculate local pressures in the presence of many-body interactions.

Related to the calculation of pressure tensors, the question of the uniqueness has been raised [4,7,14]. This discussion mainly elaborates on the possibility to add any quantity of zero divergence to the hydrodynamic pressure tensor [12]. We will not pursue this question of non-uniqueness further, which seems not to affect the physical interpretation of the pressure tensor as a force across a unit area [12]. Our treatment of local pressure tensors is entirely based on the micromechanical definition of the pressure tensor as a force across a unit area and independent from hydrodynamic theory. The results are then compared to those from hydrodynamic theory and show consistency.

The outline of the paper is as follows. In section 2, we state our definition of the local pressure tensor and the conditions for the system. We also summarize existing methods and their limitations to calculate local and average pressures. In section 3, we introduce the new method to calculate local pressures in the presence of $n$-body interactions ( $n$ = 2, 3, 4, … ). We then use this method to compare with the method of Irving-Kirkwood, the method of planes [15], and the virial theorem. In section 4, we discuss propane molecules in a cubic box as a numerical example to illustrate the application of our method in the calculation of local and average pressures, including a full comparison to results from hydrodynamics-derived methods. In section 5, we conclude the paper with a summary.

## 2. DEFINITION OF THE PRESSURE TENSOR AND EXISTING METHODS OF CALCULATION

### 2.A. Micromechanical definition of the pressure tensor

We consider a system of $N$ particles (atoms, molecules) in mechanical equilibrium with no resulting force on and no resulting velocity of its center of mass. An extension to nonequilibrium systems is feasible by subtracting local streaming velocities from the equilibrium velocities considered here. The distribution of the $N$ particles and type of interactions between them can be freely chosen, i. e., we assume $n$-body interactions ($n = 2, 3, 4, \ldots$) as the general case.

The local pressure is defined using a (infinitesimally) small cube bounded by surfaces $A_\alpha$ perpendicular to the Cartesian axes $\alpha = x, y, z$ (Fig. 1) [16]. The resulting forces $F_\beta$ in the direction of the Cartesian axes $\beta = x, y, z$ acting on each boundary area $A_\alpha$ yield the elements $p_{\beta\alpha}$ of the local pressure tensor as:

$$p_{\beta\alpha} = \frac{F_\beta}{A_\alpha}. \tag{1}$$

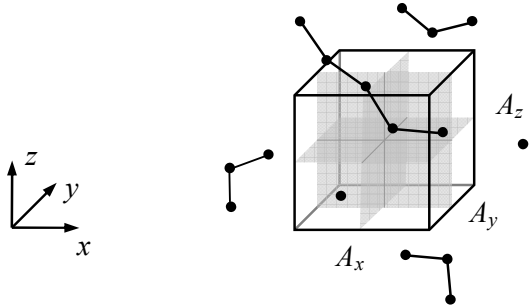

FIG. 1. Model of a small cube to illustrate the definition of local pressures. The three shaded faces $A_\alpha$ share a point of intersection in the geometric center of the cube. Some particles and molecules are also shown.

In a computer simulation, this definition of the pressure for a certain point in space (Fig. 1) is somewhat approximate because averaging over a finite volume (limit $\Delta V \to 0$) and over a certain simulation time or number of configurations (limit $\Delta t \to \infty$), respectively, is usually required. The instantaneous pressure tensor is defined by the total forces $F_x$, $F_y$, $F_z$ acting on the areas $A_x$, $A_y$, $A_z$, and there are commonly two additive contributions [17]:

$$p_{\beta\alpha} = p_{\beta\alpha}^{kin} + p_{\beta\alpha}^{int}. \tag{2}$$

These are a kinetic contribution $p_{\beta\alpha}^{kin}$ from throughput of linear momentum resulting from the particle velocities and an internal contribution $p_{\beta\alpha}^{int}$ from intermolecular and intramolecular forces acting across $A_\alpha$.

The appropriate size of the cube depends on the purpose of the calculation. The shorter the range of interactions and the more configurations are considered for averaging, the smaller may be the size of the cube. When the side length of the face $A_\alpha$ is equal or longer than the range of interactions between the particles, interactions are properly accounted for and time averaging for a short period is sufficient. If the side length of the face $A_\alpha$ is shorter than the range of interactions between the particles, the nature of these interactions may not be accounted for correctly and time averaging over a long period is recommended to avoid strong fluctuations.

**2.B. Thermodynamic definition of the pressure**

An alternative approach to calculate the pressure is based on the thermodynamic relation for fluids

$$p = -\left(\frac{\partial A}{\partial V}\right)_T, \tag{3}$$

where $A$ is the free energy of the system, $V$ the volume, and $T$ the temperature. At constant temperature, the infinitesimal change in free energy $dA$ is equal to the infinitesimal work $\delta W$ (according to the definition of $A$ and the first law of thermodynamics). In solids, $\delta W$ may be the result of tensile or shear stress, and the quantity $dV$ can be written as

$$dV_{\beta\alpha} = A_\alpha d\beta, \tag{4}$$

which represents an infinitesimal movement of the face $A_\alpha$ along the coordinate $\beta$. According to the definition of strain and shear [16], Eq. (3) becomes equivalent to Eq. (1):

$$p_{\beta\alpha} = -\left(\frac{\delta W}{dV_{\beta\alpha}}\right)_T = -\frac{1}{A_\alpha}\left(\frac{\delta W}{d\beta}\right)_T = \frac{F_\beta}{A_\alpha}. \qquad (5)$$

The minus sign indicates that the system performs work against an outside pressure, which is opposite equal to the inner equilibrium pressure (against the wall). Thus, at a microscopic level, the thermodynamic definition of the pressure tensor is equivalent to the mechanical definition in Eq. (1). Ultimately, the mechanical definition seems better suited in a simulation because difficulties to measure local free energies (or entropies) can be avoided.

**2.C. Existing methods to calculate pressure tensors**

The most important approaches are the method of Irving and Kirkwood [12], the method of planes [15], and the virial theorem [17].

*1. Local pressure tensor in the presence of two-body interactions*

Irving and Kirkwood developed a method to calculate local pressure tensors for systems with two-body interactions, based on the equations of hydrodynamics [12]. For the exact formalism, the reader is referred to the original reference [12,18]. When we consider discrete particles in a computer simulation, the mathematically exact point function pressure tensors [12] need to be extended to small cubes as introduced in section 2.A (Fig. 1). Assuming that the cubes are bounded by surfaces $A_\alpha$ with side lengths $2\Delta\alpha$, the local pressure according to Irving and Kirkwood consists of the aforementioned kinetic and internal contributions:

$$p_{\beta\alpha}(x,y,z) = \frac{1}{2A_\alpha\Delta\alpha}\left\langle\sum_{i\in(\Delta V=2A_\alpha\Delta\alpha)} m_i v_{i\beta} v_{i\alpha}\right\rangle + \frac{1}{A_\alpha}\left\langle\sum_{\vec{r}_{ij}\cap A_\alpha} F_{ij\beta}\right\rangle. \qquad (6)$$

The local pressure $p_{\beta\alpha}(x,y,z)$ at a certain coordinate $(x,y,z)$ in space is given by the masses $m_i$ and velocities $v_i$ of the particles $i$ in a small cube of volume $\Delta V = 2A_\alpha\Delta\alpha$ and by the two-body forces $\vec{F}_{ij}$ between particles whose connecting vector $\vec{r}_{ij}$ dissects the small area $A_\alpha$ (sign convention: $\vec{F}_{ij}$ is the force on the particle with higher $\alpha$ coordinate). The exact Irving-Kirkwood result is obtained for $\Delta\alpha \to 0$ ($\Delta V \to 0$) and $\Delta t \to \infty$, describing a point-function stress tensor as a result of time-averaging only [19].

Related to the Irving-Kirkwood approach are the alternative methods by Harasima [20], which defines dissection in a different way [4, 7], and the so-called IK1-approach [15], where deviations to the Irving-Kirkwood results have been reported [7].

*2. Average cross-sectional pressure*

Todd, Evans, and Daivis introduced a method to calculate average cross-sectional pressures [15]. The area $A_\alpha$ (perpendicular to axis $\alpha$) is now considered to be a cross-section of the entire box and the pressure for a given coordinate $\alpha$ is:

$$\bar{p}_{\beta\alpha}(\alpha) = \frac{1}{2A_\alpha \Delta\alpha} \left\langle \sum_{\alpha-\Delta\alpha \leq \alpha_i \leq \alpha+\Delta\alpha} m_i v_{i\beta} v_{i\alpha} \right\rangle + \frac{1}{2A_\alpha} \left\langle \sum_{i=1}^{N} F_{i\beta} \, \text{sgn}(\alpha_i - \alpha) \right\rangle. \tag{7}$$

The elements $p_{\beta\alpha}$ ($p_{x\alpha}, p_{y\alpha}, p_{z\alpha}$) of the average cross-sectional pressure at a certain coordinate $\alpha$ are given by the masses $m_i$ and velocities $v_i$ of the particles $i$ in a small cuboid of the volume $\Delta V = 2A_\alpha \Delta\alpha$ and by the net forces $\vec{F}_i$ on every particle $i$. $\vec{F}_i$ specifies the resulting force on particle $i$ due to interactions with all other particles. The exact result is formulated for $\Delta\alpha \to 0$ [15] while for computational purposes a finitely small $\Delta\alpha$ is required to sample velocities. There is no demand on the type of interactions between the particles (except for $\sum_{i=1}^{N} \vec{F}_i = 0$) so that $n$-body interactions ($n = 2, 3, 4, \ldots$) between the particles are acceptable.

*3. Average pressure over the entire box*

The virial theorem is routinely used to compute average pressures over the entire volume of a closed box [12,16,17]:

$$\bar{p}_{\beta\alpha} = \frac{1}{V} \left( \sum_{i=1}^{N} m_i v_{i\beta} v_{i\alpha} + \sum_{i=1}^{N} F_{i\beta} \, \alpha_i \right). \tag{8}$$

The elements of the average pressure tensor for the entire box are given by the masses $m_i$ and velocities $v_i$ of the particles $i$ in the total volume $V$ and by the net forces $\vec{F}_i$ on every particle $i$. The same formula rewritten for the average tensor $\bar{\bar{P}}$ reads as

$$\overline{\overline{P}} = \frac{1}{V}\left(\left\langle \sum_{i=1}^{N} m_i \vec{v}_i \cdot \vec{v}_i^T \right\rangle + \left\langle \sum_{i=1}^{N} \vec{F}_i \cdot \vec{r}_i^T \right\rangle\right) \tag{9}$$

and the scalar pressure is

$$p = \frac{p_{xx} + p_{yy} + p_{zz}}{3} = \frac{1}{3V}\left(\left\langle \sum_{i=1}^{N} m_i \vec{v}_i \cdot \vec{v}_i \right\rangle + \left\langle \sum_{i=1}^{N} \vec{F}_i \cdot \vec{r}_i \right\rangle\right). \tag{10}$$

Derivations for these formulas can be found, for example, in references [17] and [21]. Alike to the method of planes, there are no specific demands on the interaction between the particles (except for $\sum_{i=1}^{N} \vec{F}_i = 0$) and $n$-body interactions ($n$ = 2, 3, 4, …) are acceptable.

## 3. THE PRESSURE TENSOR IN THE PRESENCE OF MANY-BODY INTERACTIONS

In this section, we propose a method to calculate local pressure tensors in systems with many-body interactions and subsequently relate this method to the existing methods mentioned in the previous section.

### 3.A. The local pressure tensor in the presence of many-body interactions

Figure 2 illustrates the idea for calculating local pressure tensors. We imagine a small cube (Fig. 1) somewhere within our system and want to know what the forces across the areas $A_\alpha$ are. In computer simulations, two-body interactions (bond stretching, van-der-Waals interaction, Coulomb interaction) and many-body interactions (angle bending, torsions, out-of-plane interactions) are usually characterized by their contributions to the potential energy.

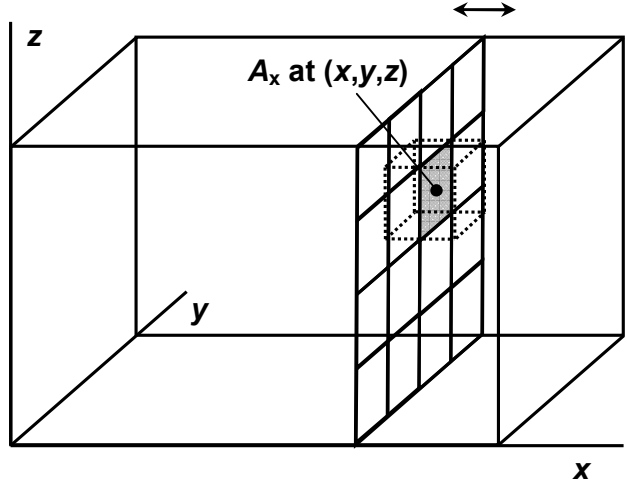

FIG. 2. Approximation of the local pressure tensor $p_{\beta\alpha}(x,y,z)$ as a function of the three coordinates. We imagine a movable grid along each of the Cartesian axes $\alpha$. For a point function pressure tensor, the microscopic cubes $\Delta V = (2\Delta\alpha)^3$ approach a zero volume.

The associated $n$-body potential $E_n$ ($n = 2, 3, 4, \ldots$) can be used to calculate a point force on every participating atom of the $n$-body interaction:

$$\vec{F}_i = -\nabla_{\vec{r}_i} E_n. \tag{11}$$

The sum over all $n$ individual forces related to the $n$-body interaction is then zero

$$\vec{F}_1 + \vec{F}_2 + \ldots + \vec{F}_n = 0 \tag{12}$$

because these interactions are internal and do not accelerate the center of mass of the associated particles. In the simplest case of two-body interactions, the two individual forces are $\vec{F}_1 = \vec{F}_{ji}$, i. e., the force exterted on particle $i$ by particle $j$, and $\vec{F}_2 = \vec{F}_{ij}$, i. e., the force exerted on particle $j$ by particle $i$. These two forces differ only in their sign and have the same absolute value, so that $\vec{F}_1 + \vec{F}_2 = 0$. Analogously, we find $\vec{F}_1 + \vec{F}_2 + \vec{F}_3 = 0$ for a three-body interaction like an angle and $\vec{F}_1 + \vec{F}_2 + \vec{F}_3 + \vec{F}_4 = 0$ for a four-body interaction like a torsion.

These interactions contribute to the pressure tensor only when they are dissected by $A_\alpha$, and a criterion for dissection must be defined. Though it is not uniquely possible, physically

9 of 27

reasonable definitions converge when the side length of the (finite) surface $A_\alpha$ is at least as large as the range of the contributing *n*-body interactions, or when sufficient time averaging is performed. We suggest that two centers of geometry are defined, for the particles of the *n*-body interaction above $A_\alpha$ and for the particles of the same *n*-body interaction below $A_\alpha$. For this purpose, we extend the finite surface $A_\alpha$ to a plane (Fig. 3). If *k* particles are located above the plane ($\alpha_i > \alpha$) and *l* particles below the plane ($\alpha_i < \alpha$), then the two centers of geometry $\vec{c}_1$ and $\vec{c}_2$ are given as the average position vector for each set of particles:

$$\vec{c}_1 = \frac{1}{k}\sum_{i=1}^{k}\vec{x}_i \; , \; \vec{c}_2 = \frac{1}{l}\sum_{i=1}^{l}\vec{x}_i \tag{13}$$

Here $\vec{x}_i$ designates the position vector of the individual particles pertaining to each of the two sets, corresponding to a total of $k + l = n$ particles for the *n*-body interaction. If the straight line $\vec{r}_n = \vec{c}_1 - \vec{c}_2$ between the two geometric centers passes through the surface $A_\alpha$, the *n*-body interaction is dissected and contributes to the internal part of the pressure. If the line between the two geometric centers does not pass through the area $A_\alpha$, no contribution to the internal part of the pressure is made (Fig. 3).

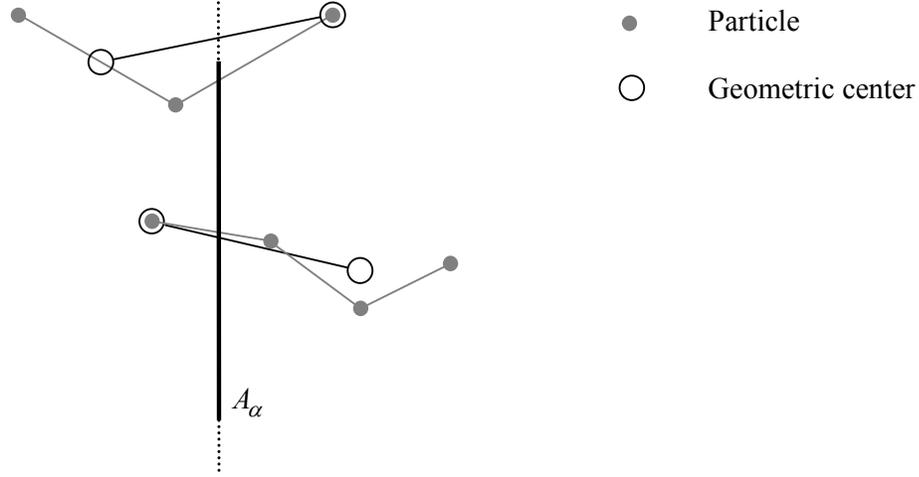

FIG. 3. The definition of geometric centers on both sides of the area $A_\alpha$, for a three-body interaction and for a four-body interaction. The connecting vector between the geometric centers of the four-body interaction is intersected by $A_\alpha$, leading to a contribution to the internal pressure. The three-body interaction does not contribute to the internal pressure across $A_\alpha$. If we assume bonds, angles, torsions, and pairwise van-der-Waals interactions (excluding 1,2 and 1,3 van-der-Waals interactions) between the particles, we find contributions from two bonds, one angle, one torsion, and 8 nonbond interactions to the internal pressure across $A_\alpha$.

Having now established the forces and a criterion of dissection, we can calculate the internal contribution to the pressure. According to Eq. (12), for each dissected *n*-body interaction the sum of point forces of the participating atoms above $A_\alpha$ is opposite equal to the sum of point forces of the participating atoms below $A_\alpha$. We can express the associated force per area $A_\alpha$ as follows:

$$p^{int}_{\beta\alpha}(x,y,z) = \frac{1}{2A_\alpha} \left\langle \sum_{\vec{r}_n \cap A_\alpha} \sum_{i=1}^{n} F_{i\beta}\, \mathrm{sgn}(\alpha_i - \alpha) \right\rangle. \tag{14}$$

The internal contribution of a *n*-body interaction to the local pressure tensor is given by the small area $A_\alpha$ and the atom-based forces $\vec{F}_i$ related to the *n*-body potential, if the connecting vector $\vec{r}_n$ between the two geometric centers on either side of $A_\alpha$ passes through $A_\alpha$. The factor ½ takes into account the effect of the sign function which counts the forces on the area $A_\alpha$ twice (the forces above and below $A_\alpha$ are opposite equal).

To obtain the complete expression for the local pressure according to Eq. (2), we add the kinetic contribution, which is the same as in Eq. (6), to Eq. (14):

$$p_{\beta\alpha}(x,y,z) = \frac{1}{2A_\alpha \Delta\alpha}\left\langle \sum_{i\in(\Delta V=2A_\alpha\Delta\alpha)} m_i v_{i\beta} v_{i\alpha} \right\rangle + \frac{1}{2A_\alpha}\left\langle \sum_{\vec{r}_n \cap A_\alpha} \sum_{i=1}^{n} F_{i\beta}\, \text{sgn}(\alpha_i - \alpha) \right\rangle. \quad (15)$$

The local pressure $p_{\beta\alpha}(x,y,z)$ at a certain coordinate $(x,y,z)$ in space is given by the masses $m_i$ and velocities $v_i$ of the particles $i$ in a small cube of volume $\Delta V = 2A_\alpha \Delta\alpha$ and by the atom-centered forces $\vec{F}_i$ due to $n$-body interactions. Only those $n$-body interactions are counted, which extend across $A_\alpha$ and whose connecting vector $\vec{r}_n$ between the two geometric centers is dissected by $A_\alpha$.

Eq. (15) is the main result of this paper. The remaining sections deal with Eq. (15) to calculate local pressures in systems with two-body interactions, average cross-sectional pressures, and volume-averaged pressures to compare with results from hydrodynamics-based approaches.

### 3.B. The special case of two-body interactions

When we consider Eq. (15) exclusively for $n=2$, the geometric centers on either side of the extended area $A_\alpha$ become identical with the two particles (Fig. 3) and we obtain $F_{1\beta} = -F_{2\beta}$ for the 2-body forces according to Eq. (12). Considering only one of the forces, we eliminate the double sum and the factor ½ in Eq. (15). When we designate the two particles as $i$ and $j$, rename their connecting vector $\vec{r}_n = \vec{r}_{ij}$, and the 2-body force $F_{1\beta} = F_{ij\beta}$, we obtain:

$$p_{\beta\alpha}(x,y,z) = \frac{1}{2A_\alpha \Delta\alpha}\left\langle \sum_{i\in(\Delta V=2A_\alpha\Delta\alpha)} m_i v_{i\beta} v_{i\alpha} \right\rangle + \frac{1}{A_\alpha}\left\langle \sum_{\vec{r}_{ij} \cap A_\alpha} F_{ij\beta} \right\rangle. \quad (16)$$

This result is the same as the Irving-Kirkwood formula for a discrete distribution of particles Eq. (6). Eq. (15) is thus consistent with the result from hydrodynamics for systems with two-body interactions.

### 3.C. The average pressure over a cross-section of the box

Averages over a cross-section of the box in the direction $\alpha$ $(x, y, z)$ are often useful to calculate cross-sectional pressure profiles. $A_\alpha$ now represents the entire cross-section of the box (Fig. 4) instead of a small area $A_\alpha = (2\Delta\alpha)^2$ within a small cube $\Delta V = (2\Delta\alpha)^3$. The kinetic contribution $\bar{p}^{kin}_{\beta\alpha}(\alpha)$ is calculated from the volume element $\Delta V = 2A_\alpha \Delta\alpha$ with a small $\Delta\alpha$. The internal contribution $\bar{p}^{int}_{\beta\alpha}(\alpha)$ is determined by the internal forces acting across the area $A_\alpha$, in the same way as a local pressure tensor. Thus, Eq. (15) is directly the solution.

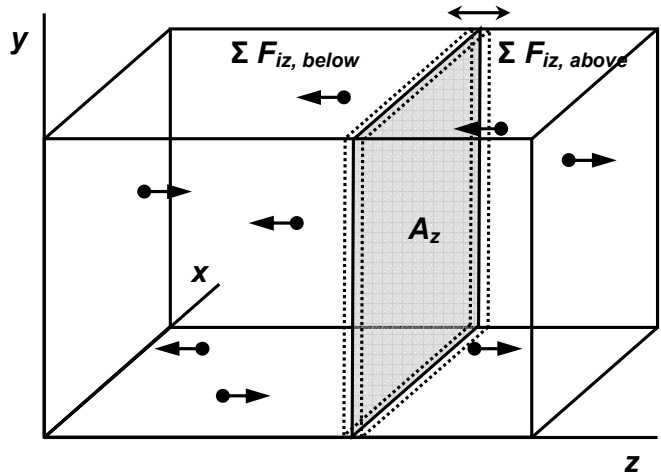

FIG. 4. Average cross-sectional pressures as a function of the coordinate $\alpha$. The calculation of the internal part of the average pressure component $\bar{p}^{int}_{zz}(z)$ is illustrated schematically.

However, simplification of this expression is possible because the cross-sectional area divides the (nonperiodic) box into two separate parts. Therefore, all *n*-body interactions with atoms on both sides of $A_\alpha$ are inevitably dissected so that we do not have to worry about the definition of geometric centers and their connecting vectors. Furthermore, all the remaining non-dissected *n*-body interactions make zero contributions (Eq. (12)) so that we can extend the double sum over all *n*-body interactions, regardless if they are dissected or not:

$$\bar{p}_{\beta\alpha}(\alpha) = \frac{1}{2A_\alpha \Delta\alpha}\left\langle \sum_{i \in (\Delta V = 2A_\alpha \Delta\alpha)} m_i v_{i\beta} v_{i\alpha} \right\rangle + \frac{1}{2A_\alpha}\left\langle \sum_{\text{all n-body interactions}} \sum_{i=1}^{n} F_{i\beta}\, \text{sgn}(\alpha_i - \alpha) \right\rangle. \quad (17)$$

The connection between individual *n*-body interactions and the net force on each atom ($\vec{F}_i = -\nabla_{\vec{r}_i} E_{pot}$) is

$$\vec{F}_i = \vec{F}_{i,n-body\ 1} + \vec{F}_{i,n-body\ 2} + \ldots + \vec{F}_{i,n-body\ k}, \quad (18)$$

i.e., the net force on each atom is constituted by the sum over all contributions from the *k* many-body interactions in which the atom is involved. Running a summation over all *n*-body interactions present in the system with their associated atom-based forces is therefore equal to running the summation over all atomic net forces:

$$\bar{p}_{\beta\alpha}(\alpha) = \frac{1}{2A_\alpha \Delta\alpha}\left\langle \sum_{i \in (\Delta V = 2A_\alpha \Delta\alpha)} m_i v_{i\beta} v_{i\alpha} \right\rangle + \frac{1}{2A_\alpha}\left\langle \sum_{i=1}^{N} F_{i\beta}\, \text{sgn}(\alpha_i - \alpha) \right\rangle. \quad (19)$$

Note that we use in Eq. (19) the symbol $F_{i\beta}$ for net forces on the atom *i* while in Equations (17) and (18) the same symbol represents the force on atom *i* due to the respective *n*-body interaction only. A simple interpretation of the result Eq. (19) is also possible in terms of the zero net force on the system as a whole ($\sum_{i=1}^{N} \vec{F}_i = 0$). The force across the plane $A_\alpha$, which divides the box in two halves, must therefore be opposite equal on both sides and be given by a summation of atom-based net forces of the particles above the plane ($\alpha_i > \alpha$), or below the plane ($\alpha_i < \alpha$), or by the summation over all forces scaled with the sign function and the factor ½ as in Eq. (19). The advantage of Eq. (19) is its simplicity: Besides coordinates and velocities, only the net forces $\vec{F}_i = -\nabla_{\vec{r}_i} E_{pot}$ acting on every particle are sufficient, which is easy to implement in molecular dynamics and Monte Carlo schemes (as long as the potential function is differentiable). Our result for the average cross-sectional pressure Eq. (19) is exactly the same as mentioned in Eq. (7) for the method of planes [15], showing consistency of our method with the result from hydrodynamics [15,22].

**3.D. The average pressure over the entire box**

Now we want to derive the average pressure over the entire box, i. e., the pressure that is felt at the box boundaries. While a derivation in the presence of 2-body interactions was given by Haile [17], we consider the presence of *n*-body interactions as the general case. We assume here again an isolated box (without periodic boundary condition) and no external forces. We view the $N$ particles in the order of increasing $\alpha$ coordinate along any of the Cartesian axes. A cross-sectional area $A_{\alpha 1}$ at $\alpha = \alpha_1$ is inserted just before the first particle is reached, a second cross-sectional area $A_{\alpha 2}$ at $\alpha = \alpha_2$ is inserted just before the second particle is reached, and so forth, until the *N*th cross-sectional area $A_{\alpha N}$ is inserted just before the *N*th particle is reached. In total, we insert $N$ imaginary planes in consecutive order along the $\alpha$ axis. The pressure due to internal forces at the plane $A_{\alpha j}$ is then given as:

$$\bar{p}_{\beta\alpha}^{\text{int}}(j) = \frac{1}{A_{\alpha j}}(F_{j\beta} + F_{j+1\beta} + \ldots + F_{N\beta}). \tag{20}$$

Here we write the net force on one side of the plane as $\sum_{i=j}^{N} \vec{F}_i$, which is equivalent to $\bar{p}_{\beta\alpha}^{\text{int}}(\alpha)$ in Eq. (19) without the factor 1/2. When we sum over all $N$ planes at their distance intervals $\Delta \alpha_i$, the average internal pressure along the entire coordinate $\alpha_0 = \sum_{i=1}^{N+1} \Delta \alpha_i$ is given as:

$$\bar{p}_{\beta\alpha}^{\text{int}} = \frac{1}{\alpha_0} \sum_{i=1}^{N+1} \bar{p}_{\beta\alpha}^{\text{int}}(i) \cdot \Delta \alpha_i . \tag{21}$$

We note that a distance $\Delta \alpha_{N+1}$ between the last particle $N$ and the end of the box must be included in the summation in order to average over the full box length $\alpha_0$. However, $\bar{p}_{\beta\alpha}^{\text{int}}(N+1) = \bar{p}_{\beta\alpha}^{\text{int}}(1) = 0$ because all particles are located on only one side of the area ($\sum_{i=1}^{N} \vec{F}_i = 0$ and Eq. (20)). Thus we can leave the $(N+1)$ term out in the summation; the first term is kept for algebraic reasons. Inserting Eq. (20) yields:

$$\bar{p}_{\beta\alpha}^{int} = \frac{1}{A_\alpha} \frac{1}{\alpha_0} \sum_{i=1}^{N} (F_{i\beta} + F_{i+1\beta} + \ldots + F_{N\beta}) \Delta\alpha_i . \tag{22}$$

Now we rearrange this expression from a sum over all $N$ planes into a sum over all $N$ particles (the particles $i$ are still labeled in the order of ascending $z$ coordinate). $\alpha_i = \Delta\alpha_1 + \ldots + \Delta\alpha_i$ is then equivalent to the $\alpha$ coordinate of the particles $i$, and together with the relation $V = A_\alpha \alpha_0$ for the box volume we obtain:

$$\bar{p}_{\beta\alpha}^{int} = \frac{1}{V} \sum_{i=1}^{N} F_{i\beta}\, \alpha_i . \tag{23}$$

This is the internal part of the pressure as in Eq. (8). The kinetic part of the average pressure over the entire box is obtained as follows: starting with the kinetic contribution of the pressure

$$\bar{p}_{\beta\alpha}^{kin}(\alpha) = \frac{1}{A_\alpha \Delta\alpha} \left\langle \sum_{\alpha - \Delta\alpha/2 \leq \alpha_i \leq \alpha + \Delta\alpha/2} m_i v_{i\beta} v_{i\alpha} \right\rangle \text{ from Eq. (19), we average over all cross-sections of}$$

an equal (infinitely) small width $\Delta\alpha$ from $\alpha = 0$ to $\alpha = \alpha_0$. We obtain

$$\bar{p}_{\beta\alpha}^{kin} = \frac{1}{V} \sum_{i=1}^{N} m_i v_{i\beta}\, v_{i\alpha} , \tag{24}$$

and the complete result according to Eq. (2) is:

$$\bar{p}_{\beta\alpha} = \frac{1}{V} \left( \sum_{i=1}^{N} m_i v_{i\beta}\, v_{i\alpha} + \sum_{i=1}^{N} F_{i\beta}\, \alpha_i \right). \tag{25}$$

This expression for the average pressure tensor over the entire box is the same as the virial formula in tensor form (Eq. (8)), showing consistency of our method with the hydrodynamic definition of the average pressure tensor over the entire box. The derivation also indicates that, in the presence of many-body interactions between the particles, the virial formula in tensor form Eq. (25) is equal to averages over cross-sectional pressures along the three Cartesian axes. This relation links the method of planes Eq. (6) or Eq. (19) [15], respectively, to the virial theorem. Equations (8), or (25), respectively, describe the average pressure tensor with respect to the walls for a rectangular box, allowing for any interactions between the particles and inhomogeneities. Besides, averaging over time is not a strict requirement and Eq. (25) is the exact pressure tensor for homogeneous systems.

In sections 3.B, 3.C, and 3.D, we have thus shown consistency of our method with known results from hydrodynamics.

### 3.E. Summary and perspectives

In conclusion, the micromechanical definition of the pressure tensor allows the calculation of local pressure tensors in the presence of many-body interactions. The method leads to the same results as existing approaches for the case of two-body interactions and average pressures.

We have also shown that the virial formula can be considered as an average over $3N$ planes along the 3 Cartesian coordinates. Important in the practical calculation of average pressures according to the method of planes Eq. (19) or the virial theorem Eq. (25) is to fulfill the condition $\sum_{i=1}^{N} \vec{F}_i = 0$. If this would not be the case, e.g., caused by approximations in the summation of Coulomb or van-der-Waals interactions, average pressures will be associated with an error on the order of

$$p_{err,\beta\alpha} = \pm \frac{1}{A_\alpha} \left\langle \sum_{i=1}^{N} F_{i\beta} \right\rangle. \tag{26}$$

## 4. NUMERICAL EXAMPLE: LIQUID PROPANE MOLECULES IN A CUBIC BOX

We consider liquid propane as a simple molecular system to analyze the influence of three-body interactions and to further validate our method. We employ a semiempirical united atom model with two types of beads for the $CH_3$ and $CH_2$ groups, as described in the NERD force field [23]. The energy expression consists of a quadratic bond stretching potential, a quadratic angle bending potential, and a 12-6 Lenard-Jones potential for van-der-Waals interactions [23].

### 4.A. Simulation details

We constructed a cubic box of $4\times4\times4$ nm$^3$ size, bounded by repulsive wall atoms on a square grid with a side length of 200 pm (Fig. 5). The wall atoms resemble carbon atoms with their effective nonbonded equilibrium distance (400 pm). Given the Lenard-Jones potential $E = A/r^{12} - B/r^6$ in the NERD force field, the parameters for the repulsive wall atoms were set to $A = 1000000$ kcal·Å$^{12}$/mol and $B = 0.01$ kcal·Å$^6$/mol.

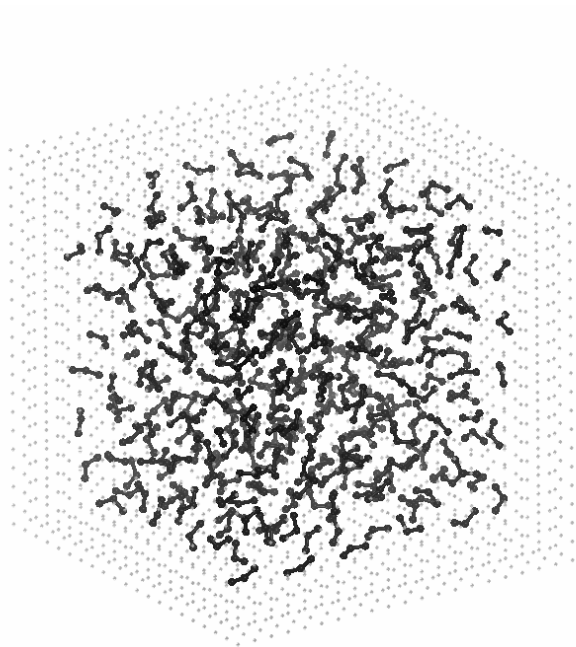

FIG. 5. Snapshot of 382 propane molecules (black) inside a 4×4×4 nm$^3$ cubic box bounded by wall atoms (gray).

The box contains 382 propane molecules, which leads to a liquid bulk density of ~520 kg/m$^3$ (see density profile further below) in agreement with the phase coexistence curve at 298 K [23]. The system was subjected to NVT dynamics, with a time step of 1 fs, the Verlet integrator, velocity scaling for temperature control, and a 1.0 nm atom-based cutoff for summation of van-der-Waals interactions. After initial equilibration of 400 ps, snapshots were recorded at intervals of one ps for analysis during a total trajectory of 400 ps, using the Discover program from Accelrys, Inc. [24].

**4.B. Relative strength of local forces by type of interaction**
Since 2-body forces and 3-body forces on each atom are the basis to calculate local pressures, we examine their relative significance first. Contributions of different interactions to the net force on each atom are displayed in Fig. 6. The graph takes into account all hydrocarbon atoms in all snapshots, i. e., ~450000 entries. It can be seen that all contributions from bond stretching, angle bending, and van-der-Waals interactions are important, although different strengths and distributions are found. Bond stretching is the strongest contribution to atom-based forces with an average of 682 pN per atom and a broad range of forces ranging from 0 pN to more than 1000 pN. This is consistent with the harmonic oscillator approximation ($F = -k_r(r(t) - r_0)$) for a continuous range of bond elongations. Similarly, angle bending is the second strongest contribution with an average of 378 pN per atom and a continuous range of forces between 0 pN to more than 600 pN. Van-der-Waals interactions account for the smallest contribution with an average of 208 pN and exhibit a narrower range of forces between 20 pN and 300 pN (Fig. 6). The bell-shaped distribution may be explained with the presence of a flexible "solvation" shell of neighbor molecules in the liquid state, which prevents atom-based forces close to zero through (dispersive) interactions and prevents excessive intermolecular forces through avoidance of close contacts.[25] The average strength of atom-based van-der-Waals forces is determined through the density of the system, e.g., an increase in density and scalar pressure would increase the average van-der-Waals force and vice versa. Fig. 6 also shows the total force on each atom, which is the vector sum over the contributions from bond stretching (2-body), angle bending (3-body) and van-der-Waals forces (2-body). The total force ranges mainly between 200 pN and 1500 pN with an average of 851 pN. This value is close to the expectation value from a "random walk" of the three constituting force vectors (809 pN), which indicates independence from each other.

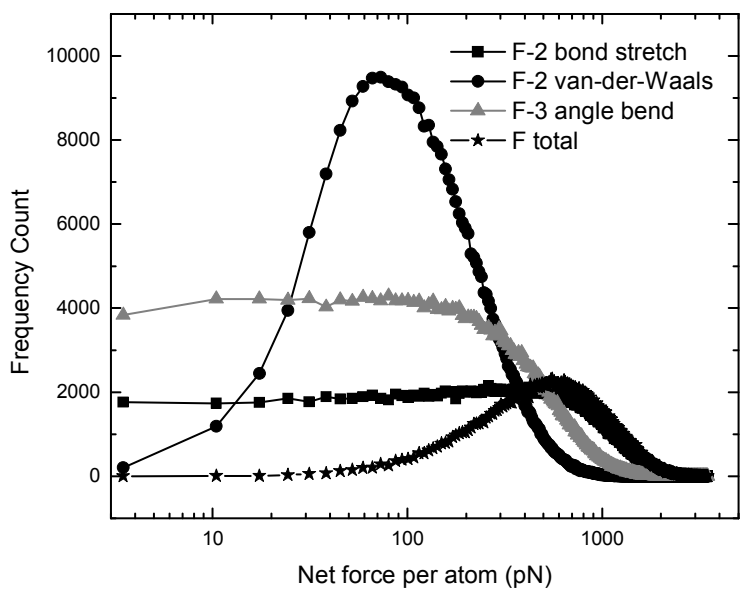

FIG. 6. Contributions to the net atomic forces and their relative significance.

**4.C. Local and average pressures**

To demonstrate the calculation of local pressures and consistency of averages with the method of planes and the virial theorem, we partition the cross-sectional area into 25 arrays, as illustrated in Fig. 7. For symmetry reasons, there are only six physically distinguishable arrays.

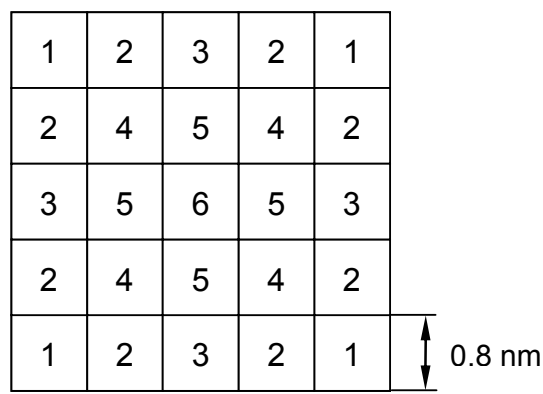

FIG. 7. Partition of the cross-sectional area of the box into arrays.

*1. Local Pressures*

The pressure profiles along the Cartesian axis perpendicular to each array are shown in Fig. 8. As a result of the cubic symmetry of the box, profiles along any of the three Cartesian axes are equal. We chose the $z$ axis and the pressure profiles for $p_{zz}(z)$ indicate 3 groups with similar pressure: array 1, arrays 2 and 3, arrays 4, 5, and 6. The increase of local pressures in this order is due to decreasing amount of "free" space between the repulsive wall and the molecules. In the center of the box, i. e., in the arrays 4, 5, 6 near $z = 2$ nm, the bulk pressure is highest (Fig. 8). Near the walls, i. e., at $z$ coordinates between 0 to 1 nm and between 3 to 4 nm, pressure fluctuations due to layering are observed (see the density profile in Fig. 8 for comparison). Local ranges and averages along the $z$ coordinate of the contributions to the pressure in array 6 are summarized in Table I. At a local scale, all contributions from two-body and three-body interactions are important. At an average (global) scale, van-der-Waals interactions (two-body) clearly dominate. This is related to the nature of these interactions to attract or repel molecules relative to each other and relative to the system boundary, thus determining the average pressure. Bonded interactions, in contrast, are internal for each molecule, contribute significantly to the local pressure, but cancel each other out as an average over a large area or a large volume. As expected for an equilibrium system, no shear pressure is found on average.

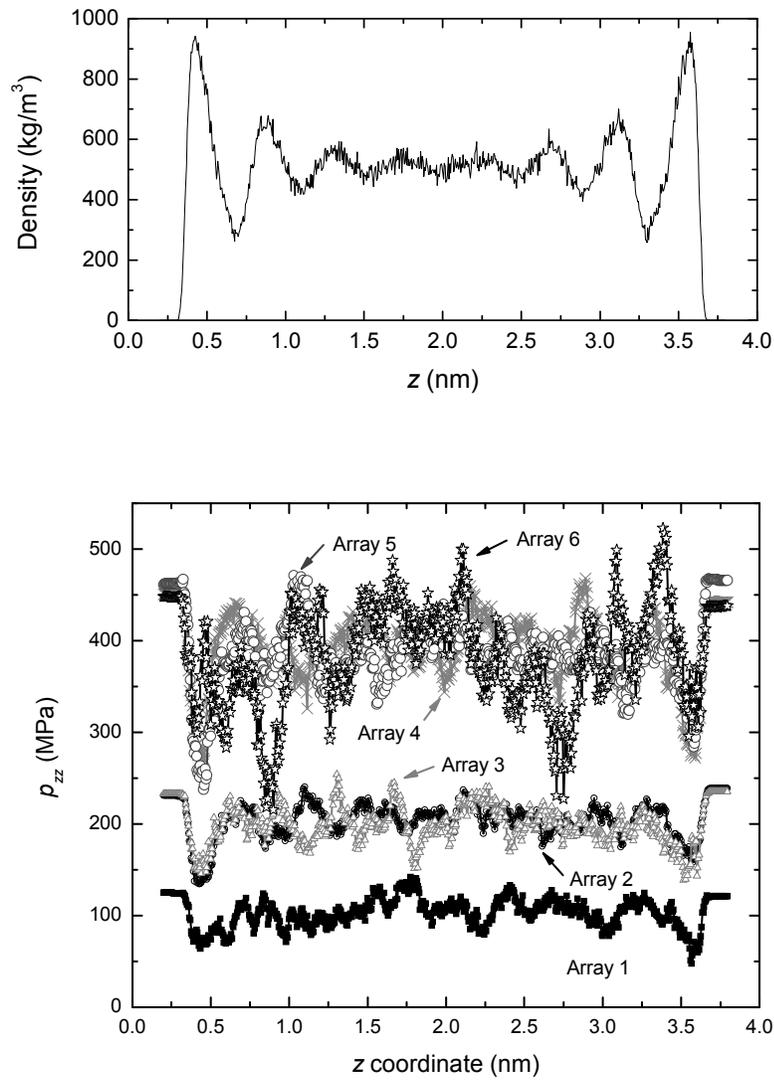

FIG. 8. Pressure profile of the $p_{zz}$ component along the $z$ axis for the arrays in Fig. 7 including kinetic, 2-body, and 3-body contributions. For reference, the density profile for the entire cross-section is also shown.

TABLE I. Range of local pressures in array 6 along the $z$ axis and average over the entire $z$ axis (MPa).

|  |  | Range | | Average |
|---|---|---|---|---|
|  |  | Lowest | Highest |  |
| $p_{xz}$ | kinetic | –25 | +25 | 0 |
|  | 2-body | –90 | +90 | 0 |
|  | 3-body | –55 | +55 | 0 |
| $p_{yz}$ | kinetic | –25 | +25 | 0 |
|  | 2-body | –90 | +90 | 0 |
|  | 3-body | –55 | +55 | 0 |
| $p_{zz}$ | kinetic | +5 | +85 | 43 |
|  | 2-body | +200 | +500 | 345 |
|  | 3-body | –55 | +55 | 0 |

*2. Average cross-sectional pressures*

We compare now the cross-sectional average of the local pressures in the 25 arrays to the method of planes. The average cross-sectional pressure is calculated by averaging over the 25 arrays and the method of planes result is independently calculated from the net atomic forces on all atoms according to Eq. (7) [15]. The fixed wall atoms are included in both calculations because they interact with the system and are needed to fulfill the condition $\sum_{i=1}^{N} \vec{F}_i = 0$. We obtain numerically identical results for cross-sectional pressures for every plane along the $z$ coordinate using both approaches. A graphical representation would therefore be little instructive (equal to a weighted average of the graphs in Fig. 8) so that we show an excerpt of the numerical results in Table II, grouped into kinetic (Kin), 2-body (I2) and 3-body (I3) contributions.

TABLE II. Cross-sectional pressures (MPa) for selected planes along the $z$ axis (nm). Averages of local pressures (Av) and results from the method of planes (MOP) are shown.

| $z$ (nm) | | $p_{xz}$ | | | $p_{yz}$ | | | $p_{zz}$ | | |
|---|---|---|---|---|---|---|---|---|---|---|
| | | Kin | I-2 | I-3 | Kin | I-2 | I-3 | Kin | I-2 | I-3 |
| 0.500 | Av | −1.216 | +16.43 | −10.68 | +0.640 | +1.992 | −0.172 | +42.63 | +171.60 | −3.421 |
| | MOP | −1.216 | +16.43 | −10.68 | +0.640 | +1.992 | −0.172 | +42.63 | +171.60 | −3.421 |
| 0.800 | Av | −0.681 | +1.832 | −2.993 | +0.472 | −1.767 | +4.228 | +34.41 | +222.25 | −0.721 |
| | MOP | −0.681 | +1.832 | −2.993 | +0.472 | −1.767 | +4.228 | +34.41 | +222.25 | −0.721 |
| 1.300 | Av | −0.230 | −4.150 | −1.528 | −0.553 | −10.80 | −0.651 | +36.68 | +225.38 | +0.441 |
| | MOP | −0.230 | −4.150 | −1.528 | −0.553 | −10.80 | −0.651 | +36.68 | +225.38 | +0.441 |
| 2.000 | Av | +0.311 | −2.626 | −0.641 | +1.139 | +2.930 | +4.855 | +31.63 | +230.33 | +4.055 |
| | MOP | +0.311 | −2.626 | −0.641 | +1.139 | +2.930 | +4.855 | +31.63 | +230.33 | +4.055 |

### 3. Average pressure over the entire box

We compare now the average pressure over the entire box calculated from local pressures to the result according to the virial formula. Local pressures of the 25 arrays were averaged to yield average cross-sectional pressures, which are now further averaged along the $z$ coordinate in small steps (section 3.D), resulting in the average pressure over the entire volume of the box. The choice of sufficiently small steps along the $z$ coordinate is important so that planes are inserted between all particles in the averaging process (here $\Delta z < 0.5$ pm). The result according to the virial theorem is independently calculated according to Eq. (25), including the fixed wall atoms. The results are shown in Table III and indicate numerical identity between the two approaches. Besides, the net atomic forces $\vec{F}_i$ can be decomposed in kinetic contributions (K), internal contributions from bond stretching (I2-B), van-der-Waals interactions (I2-V), angle bending (I3), and the total (TOT). This allows to analyze the average pressure over the entire box (as well as local pressures) with respect to all contributions (Table III). Overall, we have shown consistency between our method to calculate local pressures and their averages with the method of planes and the virial theorem.

TABLE III. Average pressure over the entire box and individual contributions (MPa). Results originating from local pressures (Av) and from the virial theorem (Vir) are identical.

| | $p_{xz}$ | | | | | $p_{yz}$ | | | | | $p_{zz}$ | | | | |
|---|---|---|---|---|---|---|---|---|---|---|---|---|---|---|---|
| | K | I2-B | I2-V | I3 | TOT | K | I2-B | I2-V | I3 | TOT | K | I2-B | I2-V | I3 | TOT |
| Av | 0.01 | 0.00 | −0.26 | −0.35 | −0.60 | −0.04 | −0.03 | −1.10 | 0.04 | −1.13 | 26.60 | 37.55 | 199.0 | −0.36 | 262.8 |
| Vir | 0.01 | 0.00 | −0.26 | −0.35 | −0.60 | −0.04 | −0.03 | −1.10 | 0.04 | −1.13 | 26.60 | 37.55 | 199.0 | −0.36 | 262.8 |

## 5. CONCLUSIONS

We consider an equilibrium system of $N$ particles with no resulting forces on and no velocity of its center of mass. Based on the mechanical definition of the pressure as a force across a unit area (Eq. (1)), we suggest a method to calculate local pressure tensors in the presence of many-body interactions (Eq. (15)). The pressure tensor contains a kinetic contribution and a contribution from dissected $n$-body interactions. If an $n$-body interaction extends across the infinite plane related to the small area, the dissection is probed by defining two geometric centers on both sides of the small area, effectively reducing the many-body interaction to a two-body interaction. For dissected $n$-body interactions, the force across the area is calculated using the point forces on each of the $n$ atoms caused by the $n$-body potential $E_n$. Consistency of the proposed method with the methods of Irving and Kirkwood, the method of planes, and the virial theorem has been shown. Besides, our approach allows to derive the method of planes and the virial theorem (an average over $3N$ planes) in a few simple steps.

As a numerical example, we analyzed the distribution of $n$-body forces for liquid propane in a cubic box. We found that both two-body forces and three-body forces are important on a local scale while only two body forces, particularly van-der-Waals forces, are important on a global (average) scale. Local pressure components have been calculated and indicate structural differences such as layering effects near the wall. The calculation of cross-sectional and volume-averaged pressures according to our method demonstrated quantitative agreement with existing approaches (method of planes, virial theorem). Local and average pressure tensors were analyzed with respect to contributions from momentum (kinetic), bond stretching, van-der-Waals interactions, and angle bending.

The possibility to calculate local pressure tensors in the presence of many-body interactions is helpful in simulations of chemically detailed systems with complex covalent bonding frameworks. Similar to the Irving-Kirkwood method, our method may also be applied under periodic boundary conditions (images need to be considered equivalent to original particles).

## Acknowledgments

Helpful discussions with Jürgen Horbach, Andrey Milchev (University of Mainz, Germany), Fatollah Varnik (Max-Planck Institute for Iron Research, Germany), and Kelly L. Anderson (Procter and Gamble Corporation, Ohio), are acknowledged. We are grateful for support by the University of Mainz, the Swiss National Science Foundation, ETH Zurich, the German National Merit Foundation, and the Air Force Research Laboratory, Wright-Patterson Air Force Base, Ohio. Comments from the referees have also been valuable.